\documentclass[prc,preprint,preprintnumbers,amsmath,amssymb,aps]{revtex4}
\usepackage{epsfig} 
\usepackage[usenames]{color}
\usepackage{subfigure}
\usepackage{graphicx}

\begin{document}

\title{ \sc Gluon condensates  in  a cold quark gluon plasma}

\author{D.A. Foga\c{c}a\dag\,  and F.S. Navarra\dag\ }
\address{\dag\ Instituto de F\'{\i}sica, Universidade de S\~{a}o Paulo\\
 C.P. 66318,  05315-970 S\~{a}o Paulo, SP, Brazil}

\vspace{3.5cm}

\begin{abstract}

The quark gluon plasma  which has been observed at RHIC is a  
strongly interacting system and has been called sQGP. This is a system at high 
temperatures and 
almost zero baryon chemical potential. A similar system with  high chemical potential and 
almost zero temperature may exist in the core of compact stars. Most likely it is also a 
strongly interacting system.  The strong interactions may be partly due to non-perturbative 
effects, which survive after the deconfinement transition and which can be related with the 
non-vanishing gluon condensates in the sQGP.  In this work, starting from the QCD Lagrangian  
we perform a gluon field decomposition in low (``soft'') and high (``hard'')  momentum components, 
we make a mean field approximation  for the hard gluons and take the matrix elements of the soft 
gluon fields in the plasma. The latter are related to the condensates of dimension two and four. 
With these approximations we derive an analytical expression for the equation of state,  which is
compared to the  MIT bag model one.  The effect of the condensates  is to soften the equation of state 
whereas the  hard gluons significantly increase the energy density and the pressure.

\end{abstract} 

\maketitle

\section{Introduction}

One of the most  interesting results of the RHIC program is the discovery of an 
extremely hot and dense state of matter made of quarks and gluons in a deconfined phase and which 
behaves like an ideal fluid \cite{shu}.  While the production of such a plasma of quarks and gluons 
had been predicted, it was a surprise to find that this system is strongly interacting and very different 
from the originally expected gas of almost non-interacting quarks and gluons, described by perturbative 
QCD.  This state has been called strongly interacting quark gluon plasma (sQGP) and there are many approaches 
to study its properties. The most fundamental approach is provided  by lattice QCD \cite{latt}. Since lattice 
QCD has still some limitations, such as the difficulty in dealing with systems with large baryon chemical 
potential, there are several models (see, for example, \cite{gelman,gardim09,bannur08})  which incorporate the essential 
features of the full theory and which can be employed to study the sQGP.  
In some of them \cite{gelman,gardim09}  the sQGP is treated as a gas of quasi-particles, 
in which the quarks and gluons have an effective mass.  In some works, such as in \cite{gelman,litim}, 
the sQGP was treated with semi-classical methods. In \cite{gelman} the color charges were assumed to be large 
and classical  obeying Wong equations of motion. In this approach the quantum effects in the QGP are basically reduced
to generate thermal-like masses and cause the effective coupling to run to larger values at smaller values of 
the  temperature.

The medium created in heavy ion collisions  has high temperature and zero baryon
chemical potential. On the other corner of the phase space, we find the QGP at zero 
temperature and high baryon number. Presumably, this kind of system exists  in the core of dense stars.  
This cold QGP has a richer phase structure and at high enough chemical potential we may have a color 
superconducting phase. Because of the limitations of lattice calculations in this domain and also because of the lack 
of experimental information, the cold QGP is less known than the hot QGP. Nevertheless it is quite possible 
that it shares some features with the hot plasma, being also a strongly interacting and semi-classical system. 

In this work we shall study the non-perturbative effects in the cold QGP generated by the residual dimension two  
and dimension four  condensates, using a mean field  approximation.

In the vacuum, non-perturbative effects have been successfully understood in terms 
of the QCD condensates, i.e., vacuum expectation values of quark and gluon  ``soft'' 
(low momentum) fields. 
The best known are the  dimension three quark condensate and the  dimension  four gluon
 condensate \cite{naricond}.
These condensates can, in principle, be computed in lattice QCD   or with the help of models. 
In practice, since they are vacuum properties and therefore universal, they can be extracted from 
phenomenological  analyses of hadron masses, as it is customary done in QCD sum rules \cite{nnl}.  The condensates are 
expected to vanish  in the limit of very high temperature or chemical potential. However, it  has been suggested  that 
they may survive after the deconfinement transition  both in the high temperature   \cite{miller07,rede}  and in the high 
chemical potential cases  \cite{zhit05}.  For our purposes the relevant gluon condensates 
are those of dimension four  \cite{naricond},  
$\big{\langle} 0 \big{|} \frac{\alpha_{s}}{\pi} {F}^{2} \big{|} 0 \big{\rangle}$ 
($ =  \langle F^2 \rangle$),  and of dimension two \cite{dudal}, 
$\big{\langle} 0 \big{|} g^2 {A}^{2} \big{|} 0 \big{\rangle}$ 
($ = \langle g^2 A^2 \rangle$).

We shall  derive an equation of state (EOS) for the cold QGP, which may be useful 
for calculations of stellar structure.  Our EOS can be considered an improved version of the EOS
of the MIT bag model, which contains both the non-perturbative effects coming from the
residual gluon condensates and the perturbative effects coming from the hard gluons, which
are enhanced by the high quark density. As it will be seen, the effect of the condensates
is to soften the EOS whereas the hard gluons significantly increase the energy density
and the pressure.

\section{The equation of state}

In this section we develop  a mean field approximation for QCD, extending previous works  
along the same line \cite{shakin,shakinn,tezuka,lovas,fukuda}. 
The Lagrangian density of QCD is given by:
\begin{equation}
{\mathcal{L}}_{QCD}=-{\frac{1}{4}}F^{a}_{\mu\nu}F^{a\mu\nu}
+\sum_{q=1}^{N_{f}}\bar{\psi}^{q}_{i}\Big[i\gamma^{\mu}(\delta_{ij}\partial_{\mu}-igT^{a}_{ij}G_{\mu}^{a})
- \delta_{ij} m_q \Big]\psi^{q}_{j}
\label{lqcdu}
\end{equation}
where
\begin{equation}
F^{a\mu\nu}=\partial^{\mu}G^{a\nu}-\partial^{\nu}G^{a\mu}+gf^{abc}G^{b\mu}G^{c\nu}
\label{efe}
\end{equation}
The summation on $q$ runs over all quark flavors, 
$m_q$ is the mass of the quark of flavor $q$,  
$i$ and $j$ are the color indices of the quarks, 
$T^{a}$ are the SU(3) generators and $f^{abc}$ are the SU(3) antisymmetric 
structure constants.  For simplicity we will consider only 
light quarks with the same mass $m$. Moreover, we will drop the summation and consider only 
one flavor. At the end of our calculation the number of flavors will be recovered.  
Following \cite{shakin, shakinn}, we shall start writting the gluon field as:
\begin{equation}
G^{a\mu}={A}^{a\mu}+{\alpha}^{a\mu}
\label{amd}
\end{equation}
where ${A}^{a\mu}$  and ${\alpha}^{a\mu}$ are the low (``soft'') and high (``hard'')  momentum
components of the gluon field respectively. The former will be responsible for the long
range and low momentum transfer, non-perturbative  processes whereas the latter will be 
relevant in the short distance perturbative processes.  The  field  decomposition made above
requires the choice  of an energy scale defining the frontier between  soft and hard. This energy 
scale, $E$, lies in the range $ \Lambda_{QCD}  <  E  <  1 \,\, \mbox{GeV}$.  In principle,  the dependence of 
the results on this choice can be studied with the renormalization group techniques.  Accurate results would  
also require the knowledge of the scale dependence  of the in-medium gluon condensates, which in our case is  
poor. Therefore, in order to keep the simplicity of our approach, we will not specify the separation scale 
and will assume that ${A}^{a\mu}$ represents the soft modes which populate the vacuum and  ${\alpha}^{a\mu}$  
represents the modes for which the running coupling constant is small. 

Inserting (\ref{amd}) into (\ref{efe}) we obtain:
$$
F^{a\mu\nu}= (\partial^{\mu}A^{a\nu}-\partial^{\nu}A^{a\mu}+gf^{abc}A^{b\mu}A^{c\nu})
+(\partial^{\mu}{\alpha}^{a\nu}-\partial^{\nu}{\alpha}^{a\mu}+gf^{abc}{\alpha}^{b\mu}{\alpha}^{c\nu}) 
$$
\begin{equation}
+ g f^{abc} {A}^{b\mu} {\alpha}^{c\nu} 
+ g f^{abc} {\alpha}^{b\mu} {A}^{c\nu} 
\label{efesplit}
\end{equation}
In the above expression the coupling is running and is  large (small) when attached to  ${A}^{a\mu}$  
(${\alpha}^{a\mu}$). The mixed terms, such as $ g f^{abc}  {\alpha}^{b\mu} {A}^{c\nu}$,  
    are assumed 
to be dominated by the large couplings.

\subsection{The mean field approximation}

In a cold quark gluon plasma the density is much larger than the ordinary nuclear matter density. 
These high densities imply a very large number of sources of the gluon field. With intense 
sources the bosonic fields tend to have large occupation numbers at all energy levels, and 
therefore they can be treated as classical fields. This is the famous approximation for 
bosonic fields used in relativistic mean field models of nuclear matter \cite{serot}. 
It has been applied to QCD in the past \cite{tezuka} and amounts to assume that the 
``hard'' gluon field, ${\alpha}_{\mu}^{a}$,  is simply a  function of the coordinates \cite{serot}:
\begin{equation}
{\alpha}_{\mu}^{a}={\alpha}_{0}^{a} \, \delta_{\mu 0}
\label{watype}
\end{equation}
In fact, for cold nuclear matter, it is further assumed that 
${\alpha}_{\mu}^{a}$ is constant in space and time \cite{serot}:
\begin{equation}
\partial_{\nu}{\alpha}^{a}_{\mu}=0
\label{delah}
\end{equation}
As a consequence of this approximation, the term $gf^{abc}{\alpha}_0^{b}{\alpha}_0^{c}$ will vanishe 
because of the color symmetry. We also assume  that the soft gluon field ${A}^{a\mu}$ is independent of position 
and time and thus:  
\begin{equation}
\partial^{\nu}{A}^{a\mu}=0
\label{dela}
\end{equation}
Substituting (\ref{watype}), (\ref{delah}) and (\ref{dela}) 
into (\ref{efesplit}) we have $F^{a\mu\nu}=g f^{abc}(A^{b\mu}A^{c\nu}+ {A}^{b\mu} {\alpha}_0^{c} \delta^{\nu 0} 
+ {\alpha}_0^{b} \delta^{\mu 0}  {A}^{c\nu})$.  Inserting this into (\ref{lqcdu}), the QCD 
Lagrangian simplifies to:
$$
{\mathcal{L}}_{QCD}=-{\frac{g^2 \, f^{abc} f^{ade}}{4}}
\Big[ {A}^{b}_{\mu} {A}^{c}_{\nu} {A}^{d \mu} {A}^{e \nu}
$$  
$$
+ {A}^{b}_{\mu} {A}^{c}_{\nu} {A}^{d \mu}  {\alpha}_0^{e}  \delta^{0 \nu}
+ {A}^{b}_{\mu} {A}^{c}_{\nu} {\alpha}_0^{d}  \delta^{0 \mu}   {A}^{e \nu}
+ {A}^{b}_{\mu} {\alpha}_0^{c} \delta_{0 \nu} {A}^{d \mu}  {A}^{e \nu}  
+ {\alpha}_0^{b} \delta_{0 \mu} {A}^{c}_{\nu}   {A}^{d \mu} {A}^{e \nu}
$$
$$
+  {A}^{b}_{\mu} {\alpha}_0^{c} \delta_{0 \nu} {\alpha}_0^{d} \delta^{0 \mu} {A}^{e \nu}
+  {A}^{b}_{\mu} {\alpha}_0^{c} \delta_{0 \nu} {A}^{d \mu}  {\alpha}_0^{e} \delta^{0 \nu} 
+  {\alpha}_0^{b} \delta_{0 \mu}  {A}^{c}_{\nu}  {A}^{d \mu}  {\alpha}_0^{e} \delta^{0 \nu}
+  {\alpha}_0^{b} \delta_{0 \mu} {A}^{c}_{\nu} \,  {\alpha}_0^{d} \delta^{0 \mu} {A}^{e \nu}
\Big]
$$
\begin{equation}
+ \bar{\psi}^{q}_{i}\Big\lbrace i \gamma^{\mu}[\delta_{ij}\partial_{\mu}-igT^{a}_{ij}
({A}_{\mu}^{a}+{\alpha}_{0}^{a}  \delta_{0 \mu})]-\delta_{ij}m\Big\rbrace\psi^{q}_{j}
\label{lqcsimp}
\end{equation}
We shall now replace the soft gluon field 
${A}^{b}_{\mu}$ and its powers by the corresponding expectation values in the 
cold QGP. The product of four fields in the first line of
the above equation can be related to the gluon condensate through the relations  
\cite{shakin,shakinn}:
\begin{equation}
\langle  {A}^{a}_{\mu} {A}^{b}_{\nu}
{A}^{c\rho} {A}^{d\eta}  \rangle
= {\frac{{\phi_{0}}^{4}}{(32)(34)}}
\Big[g_{\mu\nu}g^{\rho\eta}\delta^{ab}\delta^{cd}+
{g_{\mu}}^{\rho}{g_{\nu}}^{\eta}\delta^{ac}\delta^{bd}+
{g_{\mu}}^{\eta}{g_{\nu}}^{\rho}\delta^{ad}\delta^{bc}\Big]
\label{gc3}
\end{equation}
and
\begin{equation}  
- \frac{1}{4} \langle  {F}^{a\mu\nu}  {F}_{a\mu\nu}  \rangle 
= - \frac{\pi^2}{g^2} \langle  \frac{\alpha_s}{\pi}  
F^{a\mu\nu}F^{a}_{\mu\nu}  \rangle = - b \phi_0^4
\label{gcc}
\end{equation} 
where the constant  $b$ is given by:
\begin{equation}
b\equiv {\frac{9}{4(34)}}g^{2}
\label{bee}
\end{equation}
In the second and fourth lines of (\ref{lqcsimp}) we have odd powers of  
${A}^{a\mu}$ which have vanishing  expectation values:
\begin{equation}
\langle  {A}^{a\mu} {A}^{b\nu} {A}^{c\rho}  \rangle = 0
\label{propevani2}
\end{equation}
\begin{equation}
\langle  {A}^{a\mu}  \rangle = 0
\label{propevani1}
\end{equation} 
In the third line of (\ref{lqcsimp}) we have the hard gluon mass terms. The expectation value
of  two soft fields reads: 
\cite{shakin,shakinn}:
\begin{equation}
\langle   {A}^{a\mu} {A}^{b\nu}  \rangle 
= -{\frac{\delta^{ab}g^{\mu \nu}}{32}}{\mu_{0}}^{2}
\label{gc1}
\end{equation}
The $\langle  g^2 A^2 \rangle$ gluon condensate alone is  not gauge invariant. 
While this might be a problem in other contexts, here it is not because it appears 
always multiplied by other powers of gluon fields, forming   gauge invariant objects.  
The   $\langle  g^2 A^2 \rangle$   condensate is associated   \cite{shakin,shakinn}  with a dynamical gluon mass: 
\begin{equation}
{m_{G}}^{2}\equiv \bigg({\frac{9}{2}}\bigg)\bigg({\frac{1}{16}}\bigg)g^{2}{\mu_{0}}^{2}
\label{massagluon}
\end{equation}

In spite of the recent progress in the field, still very little is known about  
$\langle A^2 \rangle$ 
at finite (and high) density. In our approach, as in \cite{vers},  we have 
$\langle  g^2 A^2 \rangle$   $< 0$ in (\ref{gc1}) so ${m_{G}}^{2}$ is positive.

Using expressions (\ref{gc3}), (\ref{propevani2}), (\ref{propevani1}), (\ref{gc1}) and 
(\ref{massagluon})   in (\ref{lqcsimp}) we arrive at the following effective Lagrangian:
\begin{equation}
{\mathcal{L}} = -b{\phi_{0}}^{4}
+ {\frac{{m_{G}}^{2}}{2}}{\alpha}_{0}^{a}{\alpha}_{0}^{a}
+ \bar{\psi}^{q}_{i}\Big(i\delta_{ij}\gamma^{\mu}\partial_{\mu}+g\gamma^{0}T^{a}_{ij}
{\alpha}_{0}^{a}-\delta_{ij}m\Big)\psi^{q}_{j}
\label{lqcdmfttz}
\end{equation} 

\subsection{Pressure and  energy density}

From the Lagrangian (\ref{lqcdmfttz})  we can derive the  equations of motion:
\begin{equation} 
{m_{G}}^{2}{\alpha}_{0}^{a}=-g\rho^{a}
\label{azeaem}
\end{equation}
\begin{equation}
\Big(i \, \delta_{ij} \, \gamma^{\mu}\partial_{\mu}+g\gamma^{0}T^{a}_{ij} {\alpha}_{0}^{a}-m\Big)\psi_j=0
\label{psiem}
\end{equation}
where  $\rho^{a}$ is the temporal component of the  color vector current  given by:
\begin{equation}
j^{a\nu}=\bar{\psi}_{i}\gamma^{\nu}T^{a}_{ij}\psi_{j}
\label{vcur}
\end{equation}
From the Lagrangain we can obtain the energy-momentum tensor and the energy density of 
the system through:
\begin{equation}
\varepsilon=<T_{00}>
\label{eps}
\end{equation}
In the present case the energy-momentum tensor is given simply by: 
\begin{equation}
{T^{\mu}}_{\nu}={\frac{\partial \mathcal{L}}{\partial(\partial_{\mu}\psi_i)}}(\partial_{\nu}\psi_i) \,
- \, {g^{\mu}}_{\nu}\mathcal{L}
\label{tensorem}
\end{equation}
and consequently:
\begin{equation}
\varepsilon={\frac{\partial \mathcal{L}}{\partial(\partial_{0}\psi_i)}}(\partial_{0}\psi_i)
-{g}_{00}\mathcal{L} 
\label{epsilon}
\end{equation}
which, with the use of (\ref{lqcdmfttz}) gives:
\begin{equation}
\varepsilon=i\bar{\psi_i}\gamma^{0}(\partial_{0}\psi_i)
-g_{00}\Big\lbrace  -b{\phi_{0}}^{4}  
+ {\frac{{m_{G}}^{2}}{2}}{\alpha}_{0}^{a}{\alpha}_{0}^{a} 
+\bar{\psi}^{q}_{i}\Big(i\delta_{ij}\gamma^{\mu}\partial_{\mu}+g\gamma^{0}T^{a}_{ij}
{\alpha}_{0}^{a}-\delta_{ij}m\Big)\psi^{q}_{j} \Big\rbrace
\label{epsilonn}
\end{equation}
Using (\ref{psiem}) in the expression above we find
\begin{equation}
\varepsilon= b{\phi_{0}}^{4} 
-{\frac{{m_{G}}^{2}}{2}}{\alpha}_{0}^{a}{\alpha}_{0}^{a}
+i\bar{\psi_i}\gamma^{0}(\partial_{0}\psi_i)
\label{epsilonnn}
\end{equation}
Multiplying  (\ref{psiem})  by  $\bar{\psi_i}$  from the left we find:
\begin{equation}
i\psi^{\dagger}_i(\partial_{0}\psi_i)=\psi^{\dagger}_i(-i\vec{\alpha}\cdot \vec{\nabla}
+\gamma^{0}m)\psi_i-g\rho^{a}{\alpha}_{0}^{a}
\label{termo}
\end{equation}
From the usual Dirac theory applied to the study of nuclear matter we have \cite{serot}:
\begin{equation}
\psi^{\dagger}_i(-i\vec{\alpha}\cdot \vec{\nabla}
+\gamma^{0}m)\psi_i =3{\frac{\gamma_{Q}}{2{\pi}^{2}}} \int_{0}^{k_{F}}
dk \hspace{0.1cm} k^{2}\sqrt{{\vec{k}}^{2}+m^{2}}
\label{emaimportanteeeer}
\end{equation}
In the last two expressions we have:
$$
\vec{\alpha}=
\left( \begin{array}{cc}                                         
0  & \vec{\sigma}  \\
\vec{\sigma}  & 0  
\end{array} \right) \ ,
\hspace{1.2cm}
\gamma^{0}=
\left( \begin{array}{cc}                                         
1  & 0  \\
0  & -1  
\end{array} \right)
$$
where $\vec{\sigma}$ are the standard $2 \times 2$ Pauli matrices, the unit entries in $\gamma^{0}$
are $2 \times 2$ unit matrices and $\gamma_{Q}$ is the quark degeneracy factor 
$\gamma_{Q} = 2 (\mbox{spin}) \times 3 (\mbox{flavor}) $. The sum over all the color states was 
already performed and resulted in the pre-factor $3$ in the expression above. 
$k_{F}$ is the Fermi momentum defined by the quark number density $\rho$:
$$
\rho=\langle N | \psi^{\dagger}_{i} \psi_{i} | N \rangle= {\frac{3}{V}}\sum_{\vec{k},\lambda}
\langle N | N \rangle={\frac{3}{V}}\sum_{\vec{k},\lambda}=3{\frac{\gamma_{Q}}{(2\pi)^{3}}}
\int d^{3}k
=3{\frac{\gamma_{Q}}{2{\pi}^{2}}} \int_{0}^{k_{F}} dk \hspace{0.1cm} k^{2}
$$
which gives:
\begin{equation}
\rho={\frac{\gamma_{Q}}{2{\pi}^{2}}}{{k_{F}}}^{3}
\label{kf}
\end{equation}
In the above expression $|N \rangle$ denotes a state with N quarks.  
Inserting (\ref{emaimportanteeeer}) into (\ref{termo}) and  then  (\ref{termo}) into 
(\ref{epsilonnn}) we find:
\begin{equation}
\varepsilon=  b{\phi_{0}}^{4}
-{\frac{{m_{G}}^{2}}{2}}{\alpha}_{0}^{a}{\alpha}_{0}^{a}-g\rho^{a}{\alpha}_{0}^{a}
+3{\frac{\gamma_{Q}}{2{\pi}^{2}}} \int_{0}^{k_{F}}
dk \hspace{0.1cm} k^{2}\sqrt{{\vec{k}}^{2}+m^{2}}
\label{epsilonnnn}
\end{equation}
Using  (\ref{azeaem}) we can eliminate the field  ${\alpha}_{0}^{a}$ in the above 
expression:
\begin{equation}
\varepsilon= b \,  {\phi_{0}}^{4} 
+\bigg({\frac{g^{2}}{2{m_{G}}^{2}}}\bigg) \ \rho^{a}\rho^{a}  
+ \, 3 \, {\frac{\gamma_{Q}}{2{\pi}^{2}}} \int_{0}^{k_{F}}
dk \hspace{0.1cm} k^{2}\sqrt{{\vec{k}}^{2}+m^{2}}
\label{epsilonnnalend}
\end{equation}
We can relate the color charge density $\rho^{a}$ and the quark number density $\rho$. To do this we shall
use the notation of \cite{gri} and  write the quark spinor as $\psi_i = \psi c_i$, where $c_i$ is a 
color vector. We have:
\begin{equation}
\rho^{a}\rho^{a}=(\bar{\psi}_{i}\gamma^{0}T^{a}_{ij}\psi_{j})(\bar{\psi}_{k}\gamma^{0}T^{a}_{kl}\psi_{l})=
({\psi}^{\dagger}_{i}T^{a}_{ij}\psi_{j})({\psi}^{\dagger}_{k}T^{a}_{kl}\psi_{l})=
({c}^{\dagger}_{i}T^{a}_{ij}c_{j}){\psi}^{\dagger}\psi({c}^{\dagger}_{k}T^{a}_{kl}c_{l}){\psi}^{\dagger}\psi
= 3\rho^{2}
\label{roqaa}
\end{equation}
where we used the relations $\psi^{\dagger} \psi = \rho$ and  
$({c}^{\dagger}_{i}T^{a}_{ij}c_{j})({c}^{\dagger}_{k}T^{a}_{kl}c_{l})=3$. 
Performing the momentum integral we arrive at the final expression for  the energy density:
$$
\varepsilon=\bigg({\frac{3g^{2}}{2{m_{G}}^{2}}}\bigg) \ \rho^{2}
+ b \phi_0^4
+3{\frac{\gamma_{Q}}{2{\pi}^{2}}} \bigg\lbrace {\frac{{k_{F}}^{3}\sqrt{{k_{F}}^{2}+m^{2}}}{4}}+
{\frac{m^{2}{k_{F}}\sqrt{{k_{F}}^{2}+m^{2}}}{8}}
$$
\begin{equation}
-{\frac{m^{4}}{8}}ln\Big[
{k_{F}}+\sqrt{{k_{F}}^{2}+m^{2}} \ \Big] + {\frac{m^{4}}{16}}ln(m^{2}) \bigg\rbrace
\label{epsi}
\end{equation}
The pressure is given by
\begin{equation}
p={\frac{1}{3}}<T_{ii}>
\label{pr}
\end{equation}
Repeating the same steps mentioned before and using: 
\begin{equation}
\psi^{\dagger}_{i}(-i\vec{\alpha}\cdot \vec{\nabla})\psi_{i}
=3{\frac{\gamma_{Q}}{(2\pi)^{3}}}\int d^{3}k \hspace{0.1cm}
\bigg\lbrace {\frac{{\vec{k}}^{2}}{\sqrt{{\vec{k}}^{2}+m^{2}}}} \bigg\rbrace=
3{\frac{\gamma_{Q}}{2\pi^{2}}}\int_{0}^{k_{F}} dk \hspace{0.1cm} k^{2}
\bigg\lbrace {\frac{{\vec{k}}^{2}}{\sqrt{{\vec{k}}^{2}+m^{2}}}} \bigg\rbrace
\label{express}
\end{equation}
we arrive at: 
\begin{equation}
p={\frac{{m_{G}}^{2}}{2}}{\alpha}_{0}^{a}{\alpha}_{0}^{a}-b{\phi_{0}}^{4}+
{\frac{\gamma_{Q}}{2\pi^{2}}}\int_{0}^{k_{F}} dk \hspace{0.1cm} k^{2}
\bigg\lbrace {\frac{{\vec{k}}^{2}}{\sqrt{{\vec{k}}^{2}+m^{2}}}} \bigg\rbrace
\label{almp}
\end{equation}
Performing the momentum integral, using (\ref{azeaem}) and the relation for $\rho^{a}$ and the quark number density $\rho$
in (\ref{almp}) we obtain the final expression for the pressure:
$$
p=\bigg({\frac{3g^{2}}{2{m_{G}}^{2}}}\bigg) \ \rho^{2}
- b \phi_0^4
+{\frac{\gamma_{Q}}{2{\pi}^{2}}} \bigg\lbrace {\frac{{k_{F}}^{3}\sqrt{{k_{F}}^{2}+m^{2}}}{4}}-
{\frac{3m^{2}{k_{F}}\sqrt{{k_{F}}^{2}+m^{2}}}{8}}
$$
\begin{equation}
+{\frac{3m^{4}}{8}}ln\Big[
{k_{F}}+\sqrt{{k_{F}}^{2}+m^{2}} \ \Big]-{\frac{3m^{4}}{16}}ln(m^{2}) \bigg\rbrace
\label{press}
\end{equation}
The speed of sound $c_{s}$ is given by:
\begin{equation}
{c_{s}}^{2}={\frac{\partial p}{\partial \varepsilon}}
\label{cs}
\end{equation}
In the expressions above, $g$ is small, since it comes always from the coupling between the hard gluons 
and the quarks.  The large coupling is contained in the constants $b$ and $m_G$.  

Both (\ref{epsi}) and (\ref{press}) have three terms.  The first term, 
proportional to $\rho^2$,  comes from the purely hard gluonic term appearing in the 
Lagrangian and from the hard gluon term appearing in the  quark equation of motion. 
The second term, proportional to $\phi_0^4$, comes exclusively from the soft gluon 
terms and 
it has opposite signs in the energy and in the pressure. This is precisely the behavior of
the bag constant term in the MIT bag model which has the same origin. The third term 
comes from the  quarks. In short, we can say the both the energy density and the pressure 
are the sum of three contributions: the hard gluons, the soft gluons and the quarks.

\section{Numerical results and discussion}

We now compare our results (\ref{epsi}), (\ref{press}) and (\ref{cs}),  with the corresponding results 
obtained with the MIT bag model for a  gas of  quarks at zero temperature \cite{serot,nos2010}:
\begin{equation}
{{\varepsilon}_0} =\bigg({\frac{9}{4}}\bigg)\pi^{2/3}{\rho_{B}}^{4/3}
+\mathcal{B}
\label{bacanatrh}
\end{equation}
and
\begin{equation}
{{p}_0} ={\frac{1}{3}}\bigg({\frac{9}{4}}\bigg)
\pi^{2/3}{\rho_{B}}^{4/3}
-\mathcal{B}
\label{prerho}
\end{equation}
and
\begin{equation}
{{c_{0}}^{2}}={\frac{\partial p}{\partial \varepsilon}}={\frac{1}{3}} 
\label{czero}
\end{equation}
We choose  $\mathcal{B} = 110 $  MeV  fm$^{-3}$, which lies in the range ($ 50 < B < 200 $ MeV  fm$^{-3}$) used in calculations 
of stellar structure \cite{baldo,sama,burg02}. For the comparison we must rewrite 
(\ref{kf}), (\ref{epsi}) and (\ref{press})  as functions  of the baryon density, which is
  $\rho_{B}={\frac{1}{3}}\rho$. 

If we neglect the gluonic terms  and choose the quark mass $m$ to be zero in (\ref{epsi}) and 
(\ref{press})  we can show that they coincide with  (\ref{bacanatrh}) and (\ref{prerho}) with 
$\mathcal{B} = 0$.  In this limit, our model reduces to the MIT bag model. 

We next consider the MIT bag model with finite  $\mathcal{B}$ and our model with massless
quarks and soft gluons but no hard gluons. This comparison is meaningful because with these
ingredients both models describe the dynamics of free quarks under the influence of  a soft
gluon background. In this case we can identify our gluonic term with the gluonic component
of the MIT bag model,  represented by the bag constant. We then obtain an expression for the
bag constant in terms of the gluon condensate:
\begin{equation}
\mathcal{B}_{QCD}= b \phi_0^4 = \langle  \frac{1}{4} F^{a\mu\nu} F^{a}_{\mu\nu}  \rangle
\label{identif}
\end{equation}
where in the last equality we have used (\ref{gcc}) and (\ref{bee}). 
The above relation has been found in previous works, such as, for example, \cite{miller07}. 
Fixing $\mathcal{B}$ and choosing a reasonable value of the coupling of the soft gluons, 
$g$,  appearing in (\ref{gcc}) we can infer the value  of the dimension four condensate, 
$<F^2>$, in the deconfined phase. For  $\mathcal{B} = 110 $  MeV  fm$^{-3}$ and $g=2.7$ 
(which would correspond to  $\alpha_s = g^2 / 4 \pi = 0.6$) we find:
\begin{equation}
\langle   {F}^{2}  \rangle \, = \,  
\langle  \frac{\alpha_{s}} {\pi}   F^{a\mu\nu} F^{a}_{\mu\nu}  \rangle \, = \, 0.0006 \,\, \mbox{GeV}^{4}
\label{condmeio}
\end{equation}

In the lack of knowledge of  the in-medium dimension two condensate, we use the 
factorization hypothesis, which, in the notation of Refs. \cite{shakin} and 
\cite{shakinn}, implies the choice  $\mu_{0}=\phi_{0}$.  As a consequence, (\ref{gc3}), 
(\ref{gcc}), (\ref{gc1}) and (\ref{massagluon}) are related and we obtain:
\begin{equation}
\langle  g^{2} A^{2}  \rangle \, = \,  
- \sqrt{ \frac{(4)(34)\pi^{2}}{9} \langle  F^{2}  \rangle }
\, = \,  -  0.3   \,\, \mbox{GeV}^{2}
\label{factortf} 
\end{equation}
which corresponds to a dynamical mass of $m_G = 290 $ GeV. This number is consistent with 
the values quoted in recent works \cite{natale,aguilar,sorella}, which lie in the range
$200 <  m_G  <  600$ MeV.  Finally, the numerical 
evaluation of (\ref{epsi}), (\ref{press}) and (\ref{cs}) requires the choice of $g$, the 
coupling of the hard gluons, and of the quark mass, $m$.  We choose them to be $g = 0.35$ 
(corresponding to  $\alpha_s = g^2 / 4 \pi = 0.01$) and $m=0.02$ GeV. 

In Fig. 1 we show the energy density, pressure and speed of sound obtained 
with (\ref{epsi}), (\ref{press}) 
and (\ref{cs})  divided by the corresponding MIT values: 
$\varepsilon_0$, $p_0$ and $c_0$.  We observe that, 
for this set of parameters, our EOS is harder than the MIT one. This can also be seen in 
the  plot of the  pressure 
as a function of the energy density, shown in Fig. 2. In  the same range of baryon 
densities, we 
have more energy, much more pressure and consequently a larger speed of sound.  
This behavior can be 
attributed to the first term of the equations (\ref{epsi}) and (\ref{press}), which 
comes from the hard gluons. 
This term is exactly the same both in  (\ref{epsi}) and (\ref{press}) and in the 
limit of high densities becomes 
dominant yielding $p \simeq  \varepsilon $ and hence $c_s \rightarrow 1$.  
Physically, this term represents the perturbative corrections to the 
MIT approach. Since the quark density is extremely large, even in the weak 
coupling regime 
(typical of the hard gluons) the field  $\alpha^a_0$ is  intense.  A similar situation 
occurs in the color glass condensate 
(CGC). In that context, a proton (or nucleus) boosted to very high energies becomes the 
source of intense gluon fields 
generated in the weak coupling regime. Also in that case semi-classical methods were 
applied to study this gluonic system.

In Fig. 3 we plot the energy density (\ref{epsi}) (upper panel) and the pressure 
(\ref{press}) (lower panel) as a function of the baryon density $\rho_{B}$.  
We take  $\rho_{B}$  always starting at $1.5$ fm$^{-3}$. 
We can observe that the quarks and hard gluons give the dominant contributions both to  
the energy and to the pressure. Looking at the pressure we see that the hard gluons give a 
repulsive contribution  whereas the soft gluon contribution is  attractive. 
It is interesting to see that our curves follow very closely those of Refs. \cite{sama} and 
\cite{burg02}, computed with slightly different versions of the MIT bag model.  

In Fig. 4 we show the EOS for different choices of the condensates, which are now treated as
independent from each other. In the upper panel we fix $\langle F^2 \rangle$ 
and vary $\langle g^2 A^2 \rangle$, starting 
from the central value $- 0.3$ GeV$^2$ and increasing its magnitude.  
In the lower panel we perform the complementar study keeping 
$\langle g^2 A^2 \rangle $ and increasing the magnitude of $\langle F^2 \rangle$. 
As it can be seen, increasing the condensates reduces the pressure and, in the case of 
$\langle g^2 A^2 \rangle$, softens the equation of state. 
This behavior could be anticipated from Eqs. 
(\ref{epsi}), (\ref{press}) and from equation of motion (\ref{azeaem}). Indeed, keeping 
fixed the coupling and the quark density, when we increase the gluon mass, the field becomes 
weaker. In a more accurate treatment, with the inclusion of spatial inhomogeneities, the 
equation of motion (\ref{azeaem}) would contain a Laplacian term and its solution would show 
a Yukawa behavior, with the mass $m_G$ controlling the screening of the field $\alpha^a_0$.

In Fig. 5 we show  the energy per particle as a function of the baryon density for 
different values of the gluon  condensates. As in the previous figure, in the   
upper panel we fix  $\langle F^2 \rangle$ and vary $\langle g^2 A^2 \rangle$.  Increasing  
$\langle g^2 A^2 \rangle$  the energy per 
particle grows slower with baryon density. The system becomes more compressible. 
In the lower panel we  keep $\langle g^2 A^2 \rangle$  fixed and increase the magnitude of  $\langle F^2 \rangle$.
Increasing  $\langle F^2 \rangle$ leads, as before, to a more compressible system but the total energy 
is now larger.  For the central values of  $\langle F^2 \rangle$  and  $\langle g^2 A^2 \rangle$  we obtain values of  
$\varepsilon/\rho_B$ which are compatible  with those found in Ref. \cite{laura10} for 
equivalent baryon densities. As it can be seen  in all curves, the energy per particle is 
always much larger than the nucleon mass (939 MeV) and hence the system  under consideration can 
decay into nuclear matter.

To summarize, we have derived an equation of state for the cold QGP, which may be useful 
for calculations of stellar structure. The derivation is simple and based on three 
assumptions: i) decomposition of the gluon field into soft and hard components; ii) 
replacement of the soft gluon fields by their expectation values 
(``in-medium condensates'') and iii) replacement of the hard gluon fields by  their 
mean-field (classical) values.  Our EOS can be considered an improved version of the EOS 
of the MIT bag model, which contains both the non-perturbative effects coming from the 
residual gluon condensates and the perturbative effects coming from the hard gluons, which 
are enhanced by the high quark density. It is reassuring to observe that our EOS has the 
correct limits, where we recover the MIT bag model results. The  parameters are the usual  
ones in QCD calculations: couplings, masses and condensates. The effect of the condensates 
is to soften the EOS whereas the hard gluons significantly increase the energy density 
and the pressure.

\begin{figure}[t]
\includegraphics[scale=0.70]{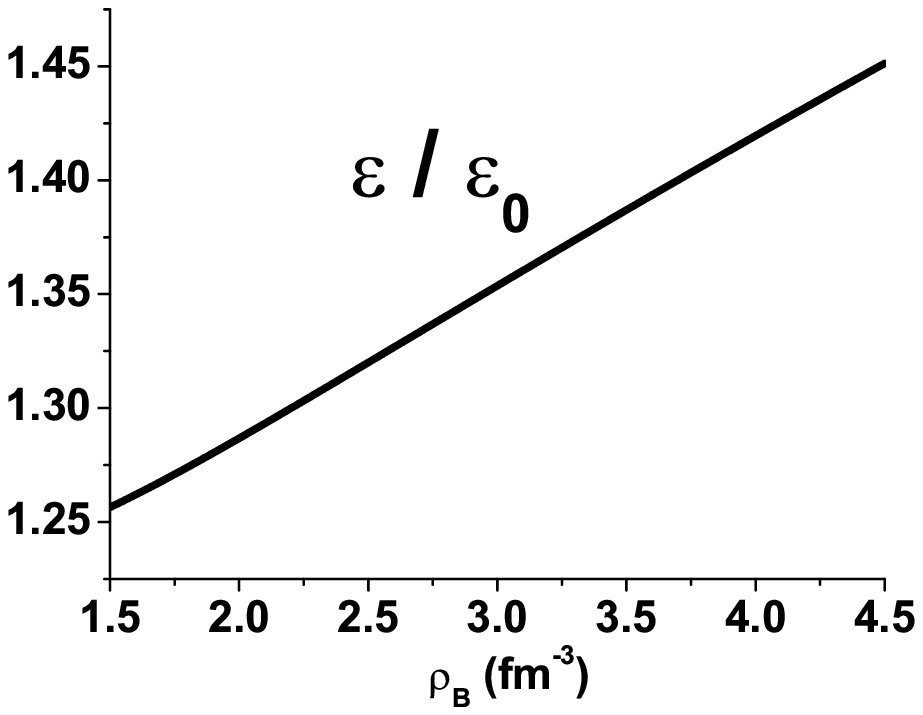}\\
\includegraphics[scale=0.70]{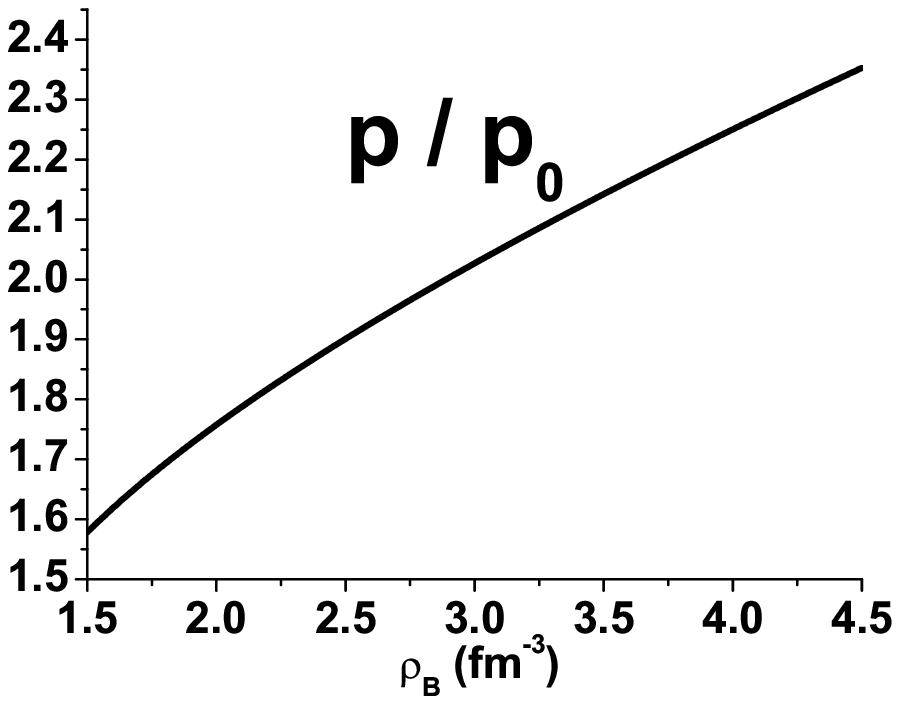}\\
\includegraphics[scale=0.70]{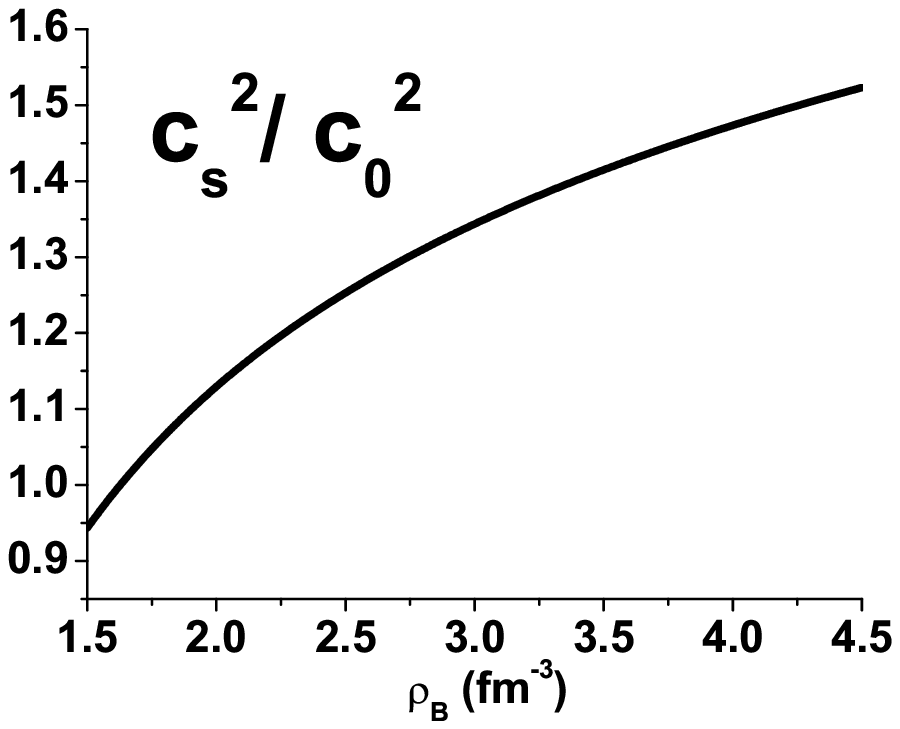}
\caption{Energy density, pressure and speed of sound, as functions of the baryon density, 
divided by the corresponding MIT values: $\varepsilon_0$, $p_0$ and $c_0$.}
\label{fig1}
\end{figure}

\begin{figure}[t]
\includegraphics[scale=0.70]{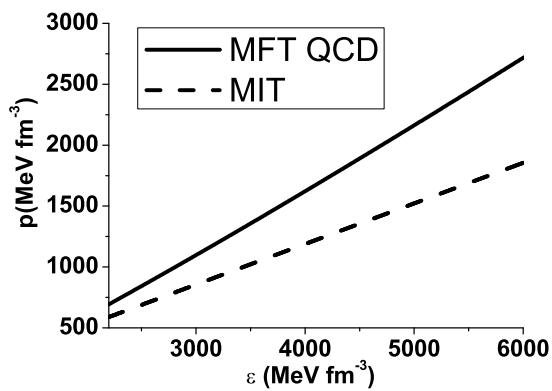}\\
\caption{Pressure as a function of the energy density.}
\label{fig2}
\end{figure}

\begin{figure}[t]
\includegraphics[scale=0.70]{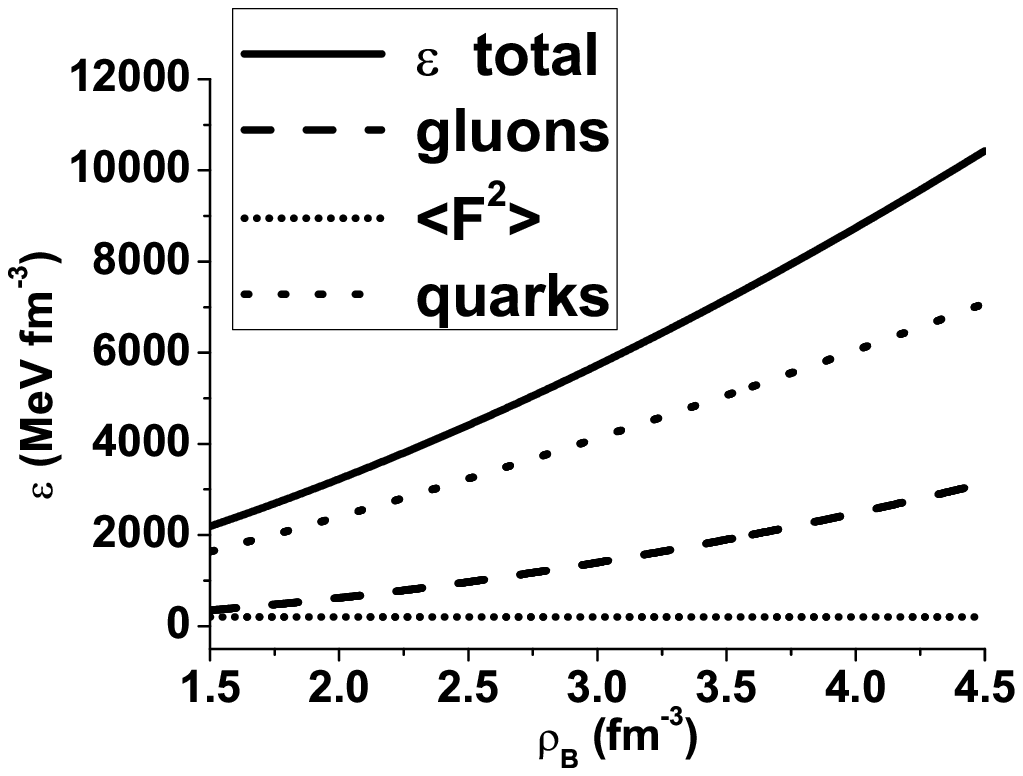}\\
\includegraphics[scale=0.70]{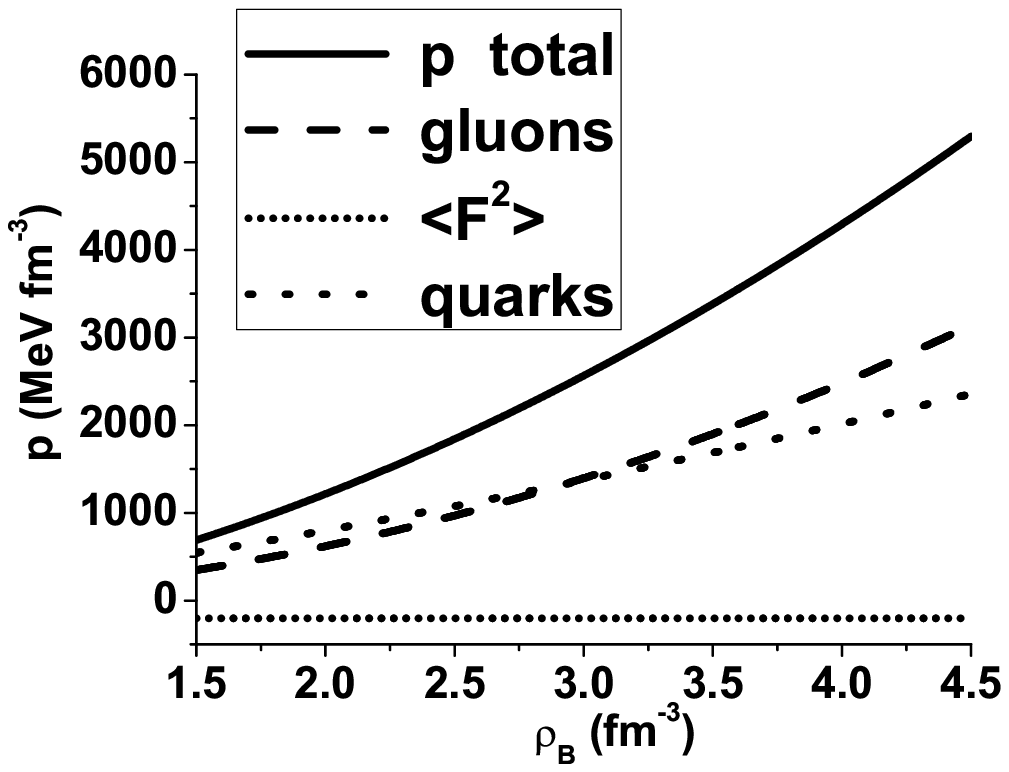}
\caption{Individual contributions to the energy density and to the pressure: 
hard gluons , quarks , soft gluons  and the sum of the three components.}
\label{fig3}
\end{figure}

\begin{figure}[t]
\includegraphics[scale=0.70]{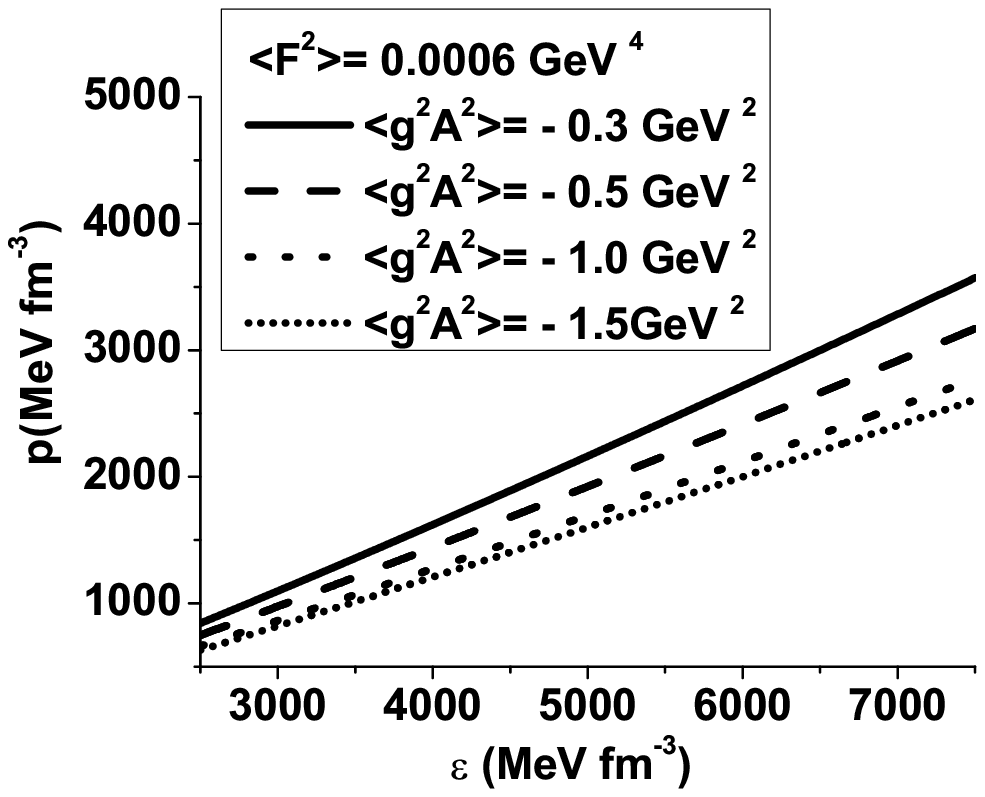}\\
\includegraphics[scale=0.70]{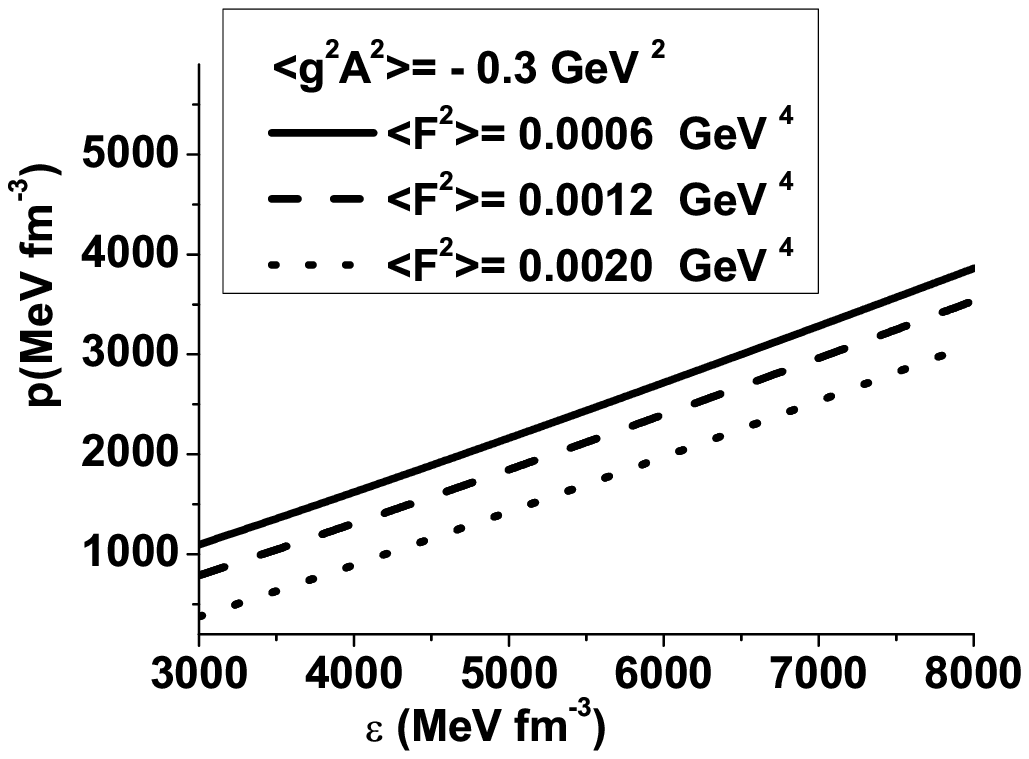}
\caption{EOS for different values of dimension two and four gluon condensates.}
\label{fig4}
\end{figure}

\begin{figure}[t]
\includegraphics[scale=0.70]{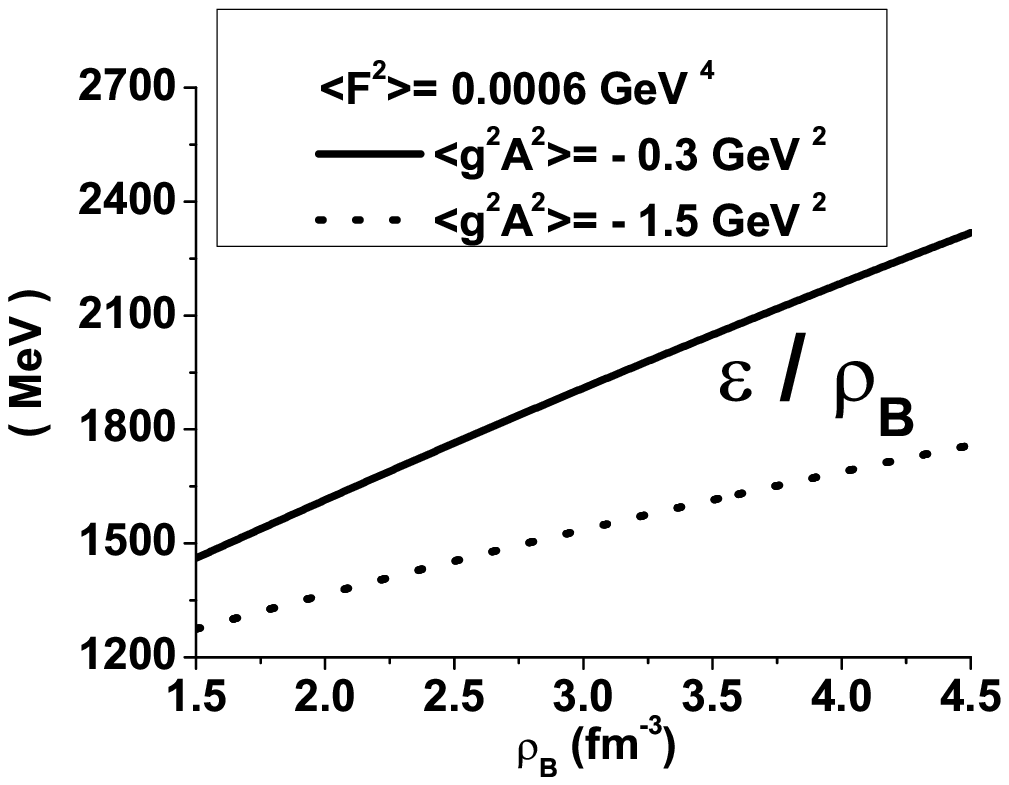}\\
\includegraphics[scale=0.70]{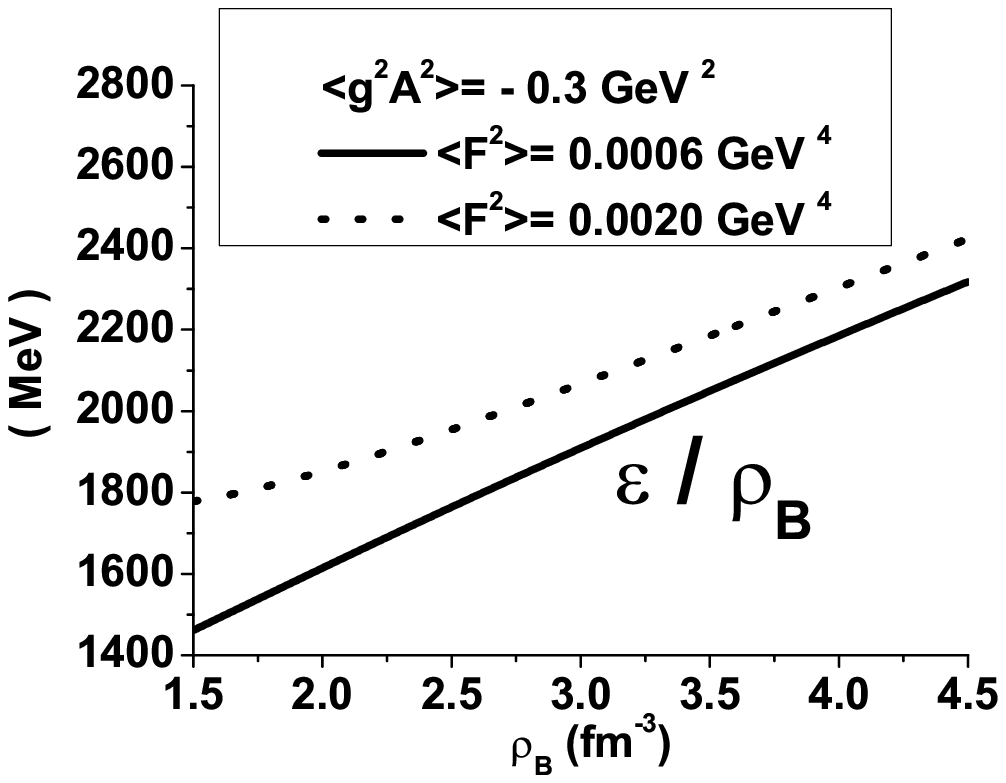}
\caption{Energy per particle as a function of the baryon density for different
values of the gluon  condensates.}
\label{fig5}
\end{figure}

\begin{acknowledgments}
We are deeply grateful to A. Natale, S. Sorella, R. Gavai and R. Gupta for fruitful 
discussions. We are especially grateful to D. Dudal for kindly and carefully answering our 
questions on the dimension two condensate. This work was  partially financed by the 
Brazilian funding agencies CAPES, CNPq and FAPESP. 
\end{acknowledgments}

\end{document}